# Geometric and statistical techniques for projective mapping of chocolate chip cookies with a large number of consumers[1]

David Orden[a]*, Encarnación Fernández-Fernández[b], Marino Tejedor-Romero[a], Alejandra Martínez-Moraian[a]

[a]Departamento de Física y Matemáticas, Universidad de Alcalá, Campus Universitario, Ctra. Madrid-Barcelona, Km. 33.600, 28805, Alcalá de Henares, Spain.

[b]Área de Tecnología de Alimentos, E.T.S. de Ingenierías Agrarias, Universidad de Valladolid, Campus La Yutera, Avda. Madrid 50, 34004, Palencia, Spain.

*Corresponding author (david.orden@uah.es)

**Abstract**

The so-called rapid sensory methods have proved to be useful for the sensory study of foods by different types of panels, from trained assessors to unexperienced consumers. Data from these methods have been traditionally analyzed using statistical techniques, with some recent works proposing the use of geometric techniques and graph theory. The present work aims to deepen this line of research introducing a new method, mixing tools from statistics and graph theory, for the analysis of data from Projective Mapping. In addition, a large number of n=349 unexperienced consumers is considered for the first time in Projective Mapping, evaluating nine commercial chocolate chips cookies which include a blind duplicate of a multinational best-selling brand and seven private labels. The data obtained are processed using the standard statistical technique Multiple Factor Analysis (MFA), the recently appeared geometric method SensoGraph using Gabriel clustering, and the novel variant introduced here which is based on the pairwise distances between samples. All methods provide the same groups of samples, with the blind duplicates appearing close together. Finally, the stability of the results is studied using bootstrapping and the RV and Mantel coefficients. The results suggest that, even for unexperienced consumers, highly stable results can be achieved for MFA and SensoGraph when considering a large enough number of assessors, around 200 for the consensus map of MFA or the global similarity matrix of SensoGraph.

---

[1] A preliminary version, entitled *Using SensoGraph for projective mapping with a large number of consumers*, was presented at Pangborn 2019.

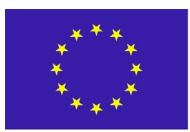

This work has received funding from the European Union's Horizon 2020 research and innovation programme under the Marie Skłodowska-Curie grant agreement No 734922.

.

# 1.-INTRODUCTION

Traditionally, the sensory evaluation of foods has been carried out using generic descriptive analysis with trained panels (Lawless & Heymann, 2010) and consumer sensory analysis with a 9-point hedonic scale, with different purposes and objectives (Varela & Ares, 2012). As an alternative to QDA and consumer sensory analysis with hedonic scale, a number of new sensory techniques have arisen in the last couple of decades (Varela & Ares, 2012; Varela & Ares, 2014; Valentin et al., 2016). These methods, sometimes called rapid sensory methods as opposed to the fact that generic descriptive analysis is quite time consuming, have proved to be useful in order to obtain accurate and reliable information from consumers.

Among these rapid methods, the present work focuses on Projective Mapping (Risvik et al., 1994), later used in Napping (Pagès, 2005). Some of the interesting characteristics of this method are being a holistic methodology, based on consumers' perception of the global similarities and differences among a set of samples, as well as allowing its use both with trained panellists and with unexperienced consumers. Data from rapid methods are traditionally analyzed using statistical techniques, with multiple factor analysis (MFA) (Pagès, 2005) being the most common choice for Projective Mapping. Very recently, some works have also proposed the use of geometric techniques from graph theory (Orden et al., 2019; Lahne, 2020). A first contribution of the present work is to deepen in this recent line of research, introducing a substantial modification of the SensoGraph method proposed in (Orden et al., 2019) and further comparing the results and their stability between this new variant, the original SensoGraph, and MFA. Our hypothesis is that mixing geometric techniques from graph theory with common tools in statistics, like distances between samples or dendrograms, would provide further insight in sensory studies.

The number of assessors required to perform Projective Mapping differs depending upon the study, the products, and the level of expertise of the participants. Previous research has shown that the minimum number of consumers needed to obtain stable maps in Projective Mapping using MFA strongly depends on the number of samples and their degree of difference (Vidal et al., 2014; Vidal et al., 2016). Although stable configurations might be reached by just 20 assessors, larger numbers are often required (Valentin et al., 2016).

Despite the simplicity of Projective Mapping makes it especially suitable to be used by unexperienced consumers, to the best of our knowledge no work seems to have carried out Projective Mapping with significantly more than 100 consumers, although other techniques like Sorting have been used with up to 389 consumers (Teillet et al., 2010). A second contribution of the present work is to perform a Projective Mapping study with a large number of consumers, in particular n=349. This large number properly allows the use of resampling techniques in order to analyze the stability of both MFA and SensoGraph for a panel of unexperienced consumers. The data collected are made publicly available (Orden & Fernández-Fernández, 2020), seeking that the community can benefit from a dataset with such a large number of consumers.

As for the product under study, Projective Mapping has been previously used to analyze a variety of foods with different sensory complexities, from white wines (Pagès, 2005; Barton et al., 2020) to cheeses (Barcenas et al., 2004; Nestrud and Lawless, 2010). For the present work we have chosen to study commercial chocolate chip cookies. Cookies are one of the most common snack foods (Gilbert et al., 2012) due to their general acceptability, convenience and shelf-life. In particular, chocolate chip cookies are present in most supermarket shelves, food stalls, and service stations. Their easy acquisition and low price help them become part of the diet, especially among young people (Ministerio de Agricultura, Pesca y Alimentación, 2019), which are the target of the present study. A Scopus search provided no previous works using Projective Mapping with chocolate chip cookies and a single one with commercial cookies (Tarrega et al., 2017). Thus, a third contribution of the present work is the analysis of commercial chocolate chip cookies, from a multinational brand and several private labels. Our hypothesis is that unexperienced consumers can provide consistent results about this popular type of cookies.

## 2.-MATERIAL AND METHODS

### 2.1 Samples

Nine commercial chocolate chip cookies were used in this study, bought at supermarkets in Palencia (Spain). Products differed in terms of brand, being or not a private label, manufacturer, and percentage of chocolate (Table 1). Blind duplicates were used within the product set, with samples 2 and 5 being the same product, since the ability of positioning close blind duplicates is widely considered an indicator of the reliability of the method and the accuracy of the panel (Moussaoui & Varela, 2010; Veinand et al., 2011; Savidan & Morris, 2015; Moelich et al., 2017).

The chocolate chip cookies were presented in plastic cups, labelled with a three-digit random code, and served in randomized order following a balanced block experimental design. Due to the significant differences in the size of the cookies and their external appearance, they were served as halves in order to minimize the possibility of the consumers recognizing them. To make sure that the assessors could re-test several times if needed, labelled plates with extra samples were at their disposal. Water was also provided to all consumers to rinse between samples.

Table 1: Chocolate chip cookies samples information.

| Sample | Brand | Private label of supermarket | Manufacturer | Percentage of Chocolate |
|---|---|---|---|---|
| 1 | Hacendado | Mercadona | Grupo Siro | 37% chocolate chips |
| 2 | Chips-Ahoy | | Mondelez España Commercial, S.L. | 25.6% chocolate chips |
| 3 | Carrefour | Carrefour | Aurly S.L. | 37% chocolate chips |
| 4 | Grandino | Lidl | Übach-Palenberg | 29% chocolate chips and 11% milk chocolate chips |
| 5 | Chips-Ahoy | | Mondelez España Commercial, S.L. | 25.6% chocolate chips |
| 6 | Alteza | Lupa | Galletas Gullón S.A. | 25% chocolate chips |
| 7 | American Cookies | Aldi | Banketbakkerij Merba B.V. | 29% chocolate chips and 11% milk chocolate chips |
| 8 | Dia | Dia | Don Cake S.A. | 26.7% chocolate chips and 10.3% milk chocolate chips |
| 9 | Ifa Eliges | Gadis | Galletas Gullón | 25% chocolate chips |

**2.2 Consumers**

A total of three hundred forty-nine (n=349) consumers participated in this study, recruited among seven educational centers and two university fairs along the academic years 2017-18 and 2018-19; more details are provided in the dataset description (Orden & Fernández-Fernández, 2020). The participants, of which 53% were female and 47% male, ranged between 14 and 30 years old. All the participants agreed to take part in this study and no identifiable or sensitive information was collected.

**2.3 Projective Mapping**

Prior to starting the tasting session, all participants were sensitized about the importance of sensory analysis by a visual presentation. An aim underlying this study was to promote the sensory analysis among young people, at different high schools and during the open days of the University of Valladolid. Subsequently, the basis of Projective Mapping was explained to the participants, using an illustration which showed cookies of different shapes (rectangular and round) and colors (cream and brown).

After the explanation of the technique, the participants received an A2 (60 x 40 cm) sheet of paper to allocate the samples. Samples were to be placed close to each other if, according to the assessor's own criteria, they seemed sensorially similar and vice versa, i.e., two chocolate chip cookies were to be distant from one another if they seemed different. The participants had to observe, smell, and taste the chocolate chip cookies, and then position the samples on the A2 sheet, trying to use as much of the tablecloth as possible. Once they had decided on the positioning, they were asked to write the codes on the sheet. The dataset with the x- and y-coordinates of chocolate chip cookies from the individual perceptual spaces is made publicly available (Orden & Fernández-Fernández, 2020), for the sake of facilitating replicability in research.

**2.4 Data analysis with existing methods**

**2.4.1 Statistical analysis**

The data obtained from Projective Mapping (Risvik et al., 1994) was then analyzed by MFA (Pagès, 2005), using the `R` language (R Development Core Team, 2007) and the `FactoMineR` package (Lê et al., 2008). Confidence ellipses were constructed using truncated total bootstrapping (Cadoret & Husson, 2013) with the `SensoMineR` package (Lê & Husson, 2008).

**2.4.2 Geometric analysis using SensoGraph with a clustering method**

The x- and y-coordinates from the tablecloths were imported and analyzed using a web app (Orden & Tejedor-Romero, 2019) implementation of the SensoGraph method introduced in (Orden et al., 2019). The first step there was to perform a clustering on each tablecloth using the Gabriel graph (Gabriel & Sokal, 1969), a tool from Computational Geometry. With this technique, two samples become connected if, and only if, there is no third sample contained in the circle having that potential connection as diameter. See Figure 1.

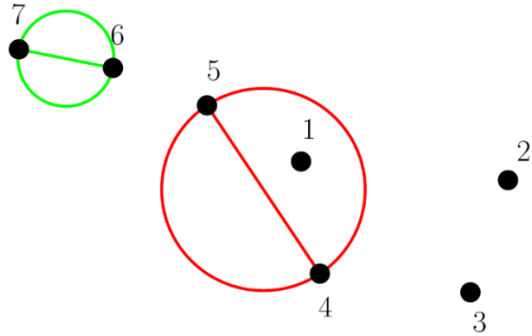

Figure 1: Clustering with the Gabriel graph. The connection 6-7 will be drawn, since no third sample lies inside the circle having that connection as diameter. On the contrary, the connection 4-5 will not be drawn, because sample 1 lies inside the corresponding circle.

The aim of this clustering was to connect some pairs of samples at each tablecloth, so that a global similarity matrix could be constructed by counting, for each pair of samples, in how many tablecloths they became connected.

In other words, for each tablecloth the clustering induced a matrix with entries in the set {0,1} standing, respectively, for the corresponding pair of samples being connected or not by the clustering. Then, the global similarity matrix was just the addition of the particular similarity matrices from each tablecloth. See Figure 2 for an illustration.

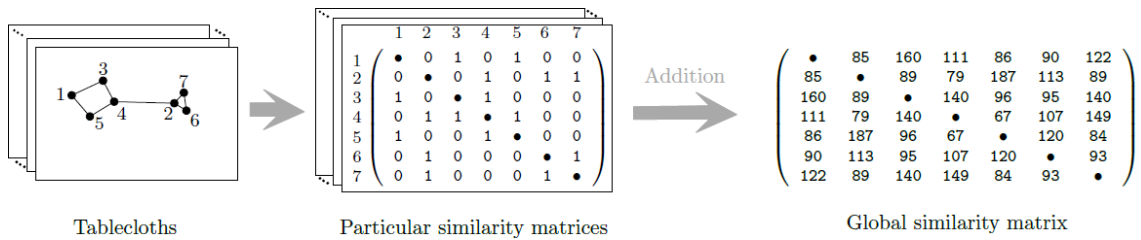

Figure 2: Illustration of the process for SensoGraph with Gabriel (Orden et al., 2019).

The second step in (Orden et al., 2019) was the use of the Kamada-Kawai force-directed algorithm from graph drawing (Kamada & Kawai, 1989) to obtain a positioning of the samples (consensus graphic) based on the entries of the global similarity matrix.

Interested readers can follow the descriptions in (Orden et al., 2019) to implement the SensoGraph method in R, using the command `gg` in the package `cccd` (Marchette, 2015) for the Gabriel graph and the command `layout_with_kk` of the package `igraph` (Csardi, 2015) for the Kamada-Kawai graph drawing algorithm. The corresponding author can also be contacted for further details.

Several improvements were implemented with respect to the software used in (Orden et al., 2019) and the features of the above-mentioned commands in R. First, a color code from red (smallest) to green (largest) was incorporated to both the connections between samples in the consensus graphic and the global similarity matrix, to illustrate the strength of those connections following the lines of recent successful visualization tools (YouGov, 2019). Second, a dendrogram for the data of the global similarity matrix was obtained using hierarchical clustering (Härdle & Simar, 2003) and the matrix was then rearranged according to the order in the dendrogram, for the sake of an easier visualization of the groups of samples.

In addition, the possibility of displaying only the most relevant connections was implemented. The strength of a particular connection was normalized to the interval between the smallest and largest strengths in the global similarity matrix, according to the following formula

$$normalized\ strength = \frac{current\ strength - smallest\ strength}{largest\ strength - smallest\ strength}$$

Then, these normalized strengths were grouped into deciles, allowing to display only the 10·k% of largest normalized strengths, for k an integer from 1 to 10.

## 2.5 Data analysis with a new method

The present work introduces and tests an alternative way of obtaining the global similarity matrix. As in the original SensoGraph proposal, the global matrix will also be the sum of particular similarity matrices for each tablecloth, but the difference is that the values in these particular matrices now lie in the interval [0,1]. See Figure 3 for an illustration.

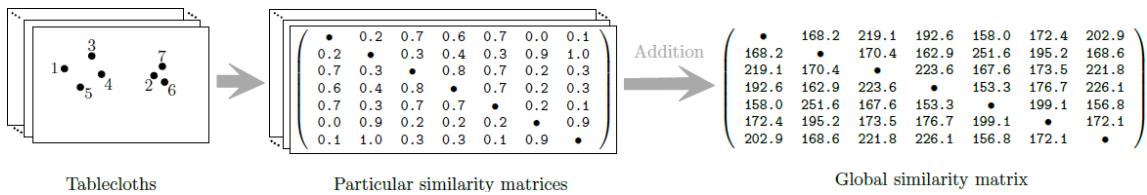

Figure 3: Illustration of the process for SensoGraph with distances.

Specifically, the values of the particular similarity matrices correspond to the distances between the pairs of samples, normalized to the interval [0,1] as follows: The interval between the smallest and the largest distances obtained is linearly mapped to a similarity in the interval [0,1] in an inverse way, according to the formula

$$similarity = 1 - \frac{distance - smallest\ distance}{largest\ distance - smallest\ distance}$$

which assigns similarity 1 to the smallest distance, 0 to the largest distance, smaller similarities (between 0 and 1) to the larger distances, and vice versa.

This approach aims to capture the essence of Projective Mapping, which asks the assessors to position closer those samples perceived as more similar and vice versa, by assigning larger similarities to (hence, considering more important) those pairs of samples positioned closer, and vice versa. In order to emphasize the importance of those samples which were more clearly positioned closer or further, the linear mapping mentioned above was tuned using the following formula, which depends on a parameter p≥1

$$tuned\ similarity = \begin{cases} 2^{p-1} \cdot (similarity)^p & ,when\ similarity\ < 1/2 \\ 1 - 2^{p-1} \cdot |similarity - 1|^p & ,when\ similarity\ \geq 1/2 \end{cases}$$

Figure 4 shows the effect of this function for several values of p. The value p=1 gives the aforementioned linear mapping, while increasing the value of p results in the graph combing towards an S-shape, hence emphasizing the effect of extreme values corresponding to more clearly positioned samples. After checking different values of p between 1 and 3 for the function of tuned similarity, obtaining analogous results, a compromise value of p=2 was chosen for the sake of emphasizing extreme values without an excessive distortion of the inputs, see again Figure 4.

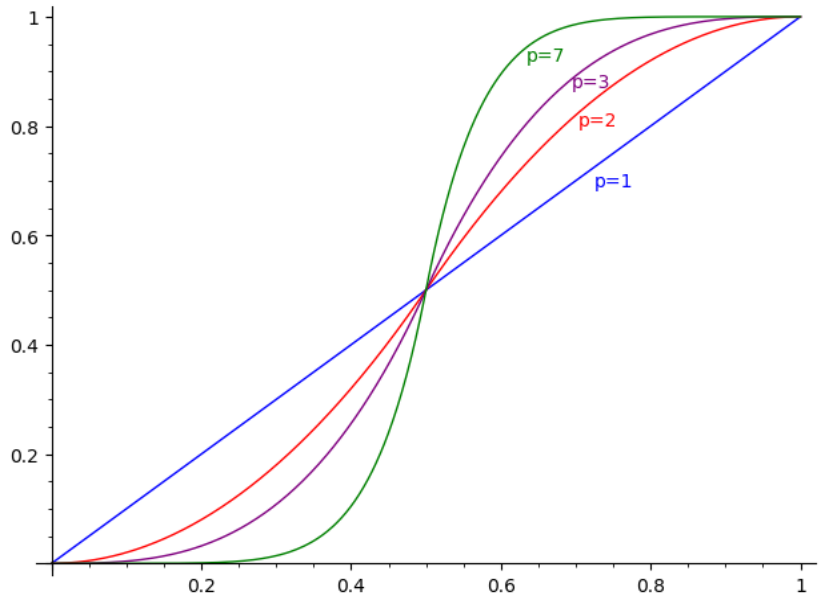

Figure 4: Effect of linear (p=1) and non-linear mappings (p>1) over the interval [0,1].

This new method was also implemented in the web app (Orden & Tejedor-Romero, 2019), using Python as backend language on the server side, Flask as microframework, and MongoDB to manage the databases, as well as JavaScript, HTML, and CSS on the user side. Again, replicability of the results can be ensured by alternative implementations using the techniques and commands detailed above. Interested readers can also contact the corresponding author for further details.

**2.6 Stability of the results**

The stability of the results obtained was analyzed using bootstrapping resampling in order to simulate repetition of the experiments (Shao & Tu, 1995), as used by Vidal et al. (2014) for the study of the stability of sample configurations from projective mapping. For each value of m=10, 20, 30,…,n, subsets of m assessors were randomly drawn with replacement from the original data set. As in previous works (Faye et al., 2006, Blancher et al., 2012), a collection of 100 subsets was generated for each value of m.

For the study of the stability of MFA results three analyses were performed, using independent bootstrapping resamplings: With the first two dimensions of the MFA and with the first four dimensions as in the stability study by Vidal et al. (2014) and, in addition, with eight dimensions for better comparison with the results for SensoGraph.

In order to measure the agreement between the MFA consensus map for each subset and that of the original panel, the RV coefficient was computed (Escoufier, 1973; Tomic et al., 2015; Josse & Holmes, 2016). This is a popular similarity measure between point configurations or matrices, whose values range between 0 and 1 and for which, the more similar two items are, the higher is the corresponding RV coefficient.

The RV coefficient was computed using the `FactoMineR` (Lê et al., 2008) function `coeffRV`. The mean and the standard deviation for RV coefficients of subsets of the same size were then obtained using the commands `mean` and `sd` of the `R` language (R Development Core Team, 2007).

For the study of the stability of SensoGraph results, the only meaningful analysis is considering the global similarity matrices. A graph is defined by a set of vertices and a set of pairs *(i,j)* of vertices (Gross et al., 2013), which are encoded as the set of entries *(i,j)* of a matrix. The two-dimensional representation of a global similarity matrix is a plotting artifact based on the graph-drawing algorithm considered. Furthermore, the random nature of the seeds chosen by the Kamada-Kawai algorithm used by SensoGraph could result in different coordinates in the consensus graph.

In order to measure the agreement between the SensoGraph similarity matrix for each subset and that of the original panel, the Mantel coefficient was computed (Mantel, 1967), which evaluates the similarity between two configurations by measuring the correlation between two matrices of distances between the samples (Abdi, 2010). The Mantel coefficient is the usual Pearson correlation coefficient based on two vectors of size n(n-1)/2 (n being the number of products) containing the off-diagonal elements of the two dissimilarity matrices being compared. Its values range between -1 and 1 and, the more similar two items are, the higher is the corresponding Mantel coefficient. Widely used in ecology, the Mantel coefficient has also been used in sensory analysis, e.g., by Blancher et al. (2012) to investigate the stability of sorting maps.

The mantel coefficient was computed using the Python function `skbio.stats.distance.mantel` (The scikit-bio development team, 2020) using Pearson's product-moment correlation coefficient. The mean and the standard deviation for Mantel coefficients of subsets of the same size were then obtained using the commands `mean dataset.mean` and `dataset.std` of the pandas library (The pandas development team, 2020).

## 3.-RESULTS AND DISCUSSION

### 3.1 Results for MFA

The data obtained from the Projective Mapping tablecloths were exported as a CSV file and were processed using MFA as mentioned in Section 2.4.1. The computation took an average of 90.2 seconds, with a standard deviation of 11.7 seconds, measured over 5 independent runs on a computer with an Intel(R) Core(TM)2 Duo CPU @2.13GHz with 4GB of RAM.

The consensus representation of the similarities and differences among samples is shown in Figure 5. In this consensus graphic of MFA, the first two dimensions accounted for 41.12% of the explained variance (25.64% Dim1 and 15.48% Dim2). Low percentages of explained variance have been observed for panels composed by unexperienced consumers (Nestrud & Lawless, 2010), with some works reporting such panels not positioning close together blind duplicate samples (Nestrud & Lawless, 2008).

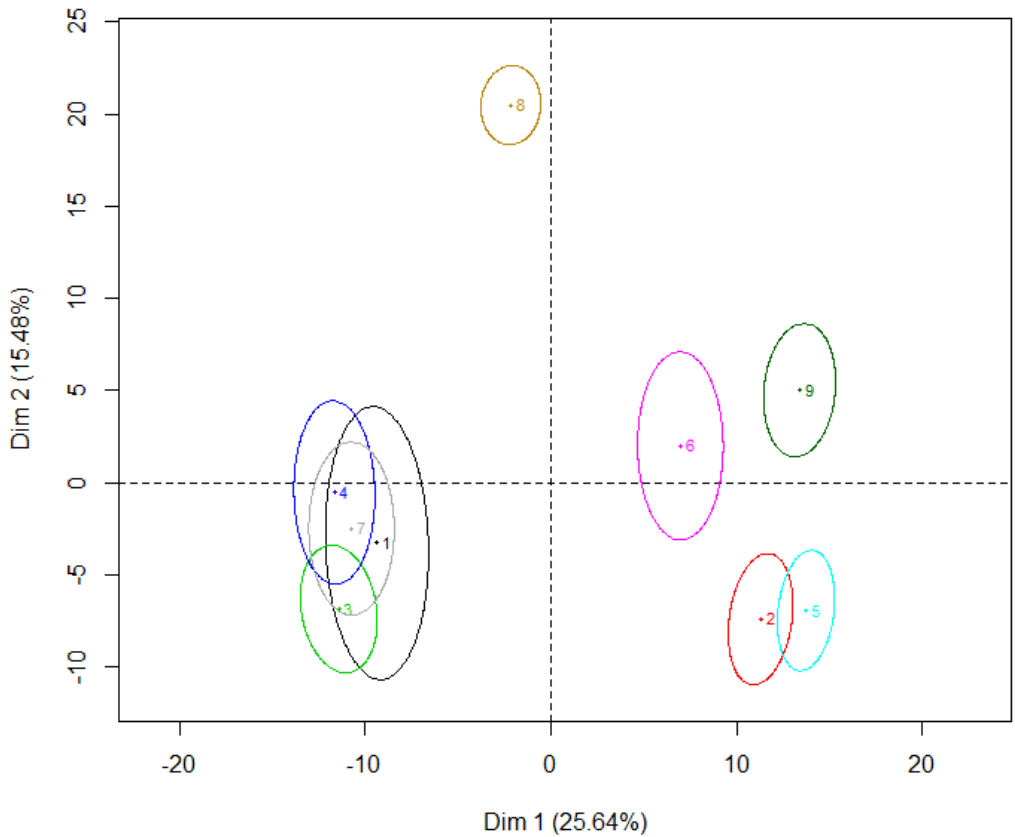

Figure 5: Consensus plot from MFA, dimensions 1 and 2.

In our case, the panel of unexperienced consumers was able to position the blind duplicate samples 2 and 5, from the same brand (Chips-Ahoy), close together in the lower-right quadrant. Their confidence ellipses overlapping means that no significant differences were perceived and, therefore, the consumers were capable to detect their similarity.

In addition to that, samples 1 (Hacendado), 3 (Carrefour), 4 (Grandino), and 7 (American Cookies) became positioned in the lower-left quadrant of the perceptual space, with their confidence ellipses overlapping indicating that consumers did not perceive statistically significant differences between them. These were the cookies with the highest percentage of chocolate: Samples 1 (Hacendado) and 3 (Carrefour) had a 37% of chocolate chips, while samples 4 (Grandino) and 7 (American Cookies) had a 29% of chocolate chips plus an 11% of milk chocolate chips. Recall Table 1.

Further, sample 6 (Alteza) and sample 9 (Ifa Eliges) appear separated, with their ellipses not overlapping but not far to each other and to samples 2 and 5. Probably, this can be explained because the samples 6 and 9 have a percentage of chocolate similar to samples 2 and 5 (25% and 25.6% of chocolate chips, respectively, see Table 1). Samples 6 and 9 are from a private label, but produced by the same manufacturer, while the duplicate samples 2 and 5 are from a best-selling multinational brand (see Table 1).

Finally, sample 8 (Dia), was isolated at the upper-left quadrant despite its proportion of chocolate chips being similar to others (26.7% of chocolate chips and 10.3% of milk chocolate chips, see Table 1). A possible explanation is that this sample appears as the farthest one to samples 2 and 5, those from a best-selling multinational brand, so the private label cookie 8 was perceived as the least similar to a best-selling cookie.

The above discussion is to be considered taking into account that, due to the lack of batch tracking, changes in product recipes might have arisen over the two years of data collection.

**3.2 Results for SensoGraph with Gabriel clustering**

The data from the Projective Mapping tablecloths was uploaded to the SensoGraph software (Orden & Tejedor-Romero, 2019), which processed them as described in Section 2.4.2. The computation took less than one second with the same computer mentioned in Section 3.1.

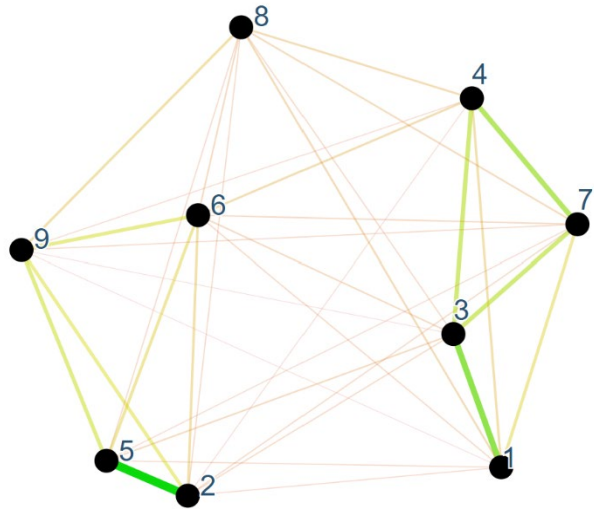

Figure 6: Consensus plot obtained using SensoGraph with Gabriel (Orden et al., 2019).

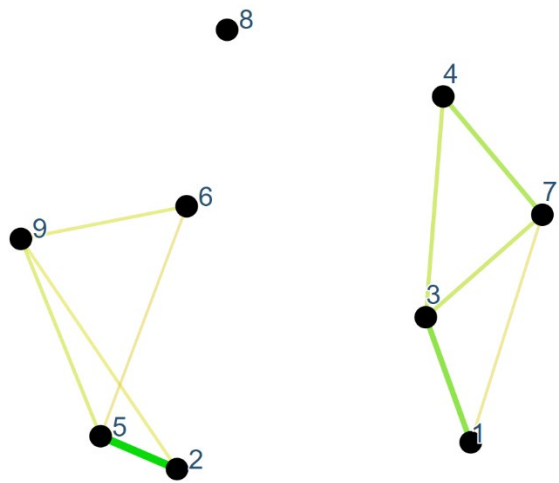

Figure 7: Consensus plot from SensoGraph with Gabriel, showing only the 60% of most relevant edges.

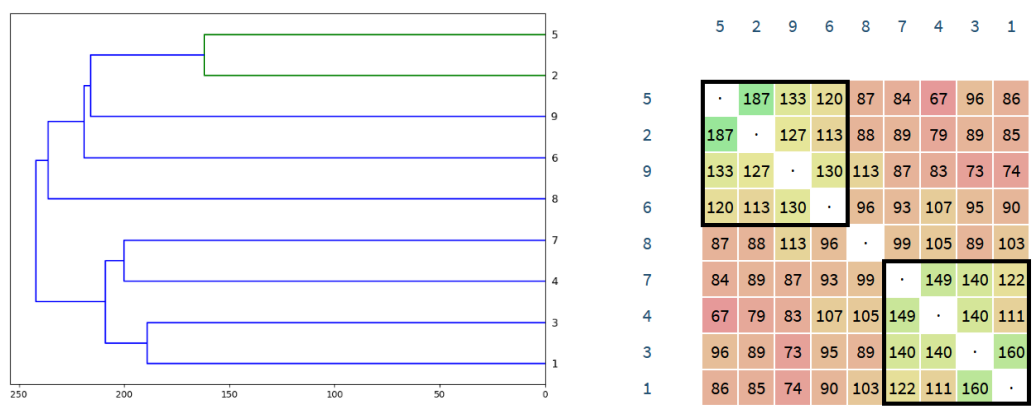

Figure 8: Dendrogram (left) and global similarity matrix (right) for SensoGraph with Gabriel. The matrix is rearranged according to the result of the dendrogram, so that the groups obtained appear as submatrices with similar colors (framed).

Figure 6 shows the consensus graphic obtained, with all the connections between samples, obtained using SensoGraph with the Gabriel graph clustering as in (Orden et al., 2019), see Section 2.4.2. Figure 7 shows only the 60% of most relevant connections, while Figure 8 shows the dendrogram (left) obtained by hierarchical clustering from the global similarity matrix with the strengths of those connections (right). For the consensus plots and the global similarity matrix, a color code from red (smallest strength) to green (largest strength) has been used for the sake of an easier visualization.

These graphics show that the positioning of the samples provided by SensoGraph is similar to that in the consensus map given by MFA, the RV coefficient between the point configurations in Figures 5 and 6 being 0.8785. Groups 2-5-6-9, 1-3-4-7, and 8 can be identified in Figures 7 and 8, with the global similarity matrix in Figure 8 (right) easily showing that the connection 2-5 is the strongest (greenest) one, corresponding to samples 2 and 5 being the blind duplicates. It is interesting to note that SensoGraph identifies these blind duplicates as the most similar samples, both by the consensus map in Figure 6 and by the global similarity matrix in Figure 8, while in the MFA graphic for the first two dimensions (Figure 5) the pair 1-7 is instead the closest one (although the confidence ellipses for 2-5 do overlap, as discussed above). Nevertheless, note that the value of the connection 2-5 in the SensoGraph similarity matrix is just 187 out of the 349 consumers considered.

In addition, SensoGraph shows samples 6 and 9 appearing in the same group as the aforementioned samples 2 and 5, although the strength of the other connections in this group is quite smaller than that of the connection 2-5. Samples in this group share a similar percentage of chocolate chips since, as commented above, samples 6 (Alteza) and 9 (Ifa Eliges) have a 25% of chocolate chips, and samples 2 and 5 (Chips-Ahoy) have a 25.6% of chocolate chips (Table 1).

A second group observed in the SensoGraph graphic is composed by samples 1-3-4-7, with the corresponding ellipses overlapping in MFA as well. The connection 1-3 is the second strongest one and the connection 4-7 is the third strongest one, while connections 1-4 and 1-7 are not so strong, and the connections 3-4 and 3-7 have an intermediate strength. This group can be explained because the samples 1 (Hacendado) and 3 (Carrefour) have the same percentage of chocolate chips, 37%, while the sample 4 (Grandino) and the sample 7 (American Cookies) also have both the same percentage of chocolate (29% of chocolate chips plus 11% of milk chocolate chips).

Finally, the sample 8 (Dia) being isolated in MFA coincides with its connections to all other samples being weak in SensoGraph, and with this sample being actually isolated when showing only the 60% of most relevant edges (Figure 7). As happened for MFA, the sample 8 appears as the farthest one to samples 2 and 5 also in SensoGraph, both for the consensus map (Figure 6) and for the global similarity matrix (Figure 8, right).

### 3.3 Results for SensoGraph with distances

In this case, the data was uploaded to the software by Orden & Tejedor-Romero (2019) and the option of using SensoGraph with distances was selected, in order to process the data as described in Section 2.5. The computation took less than one second with the same computer mentioned in Section 3.1.

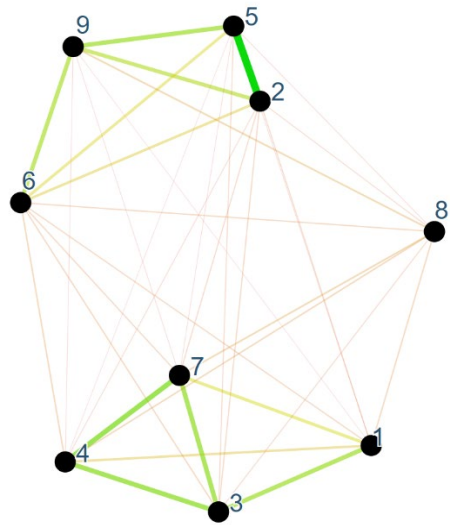

Figure 9: Consensus plot obtained using SensoGraph with distances, as introduced in Section 2.5.

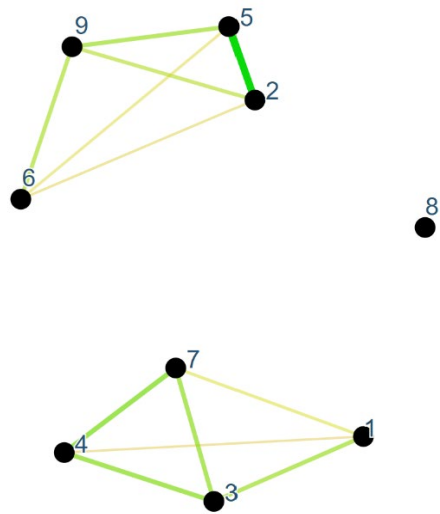

Figure 10: Consensus plot from SensoGraph with distances, showing only the 60% of most relevant edges.

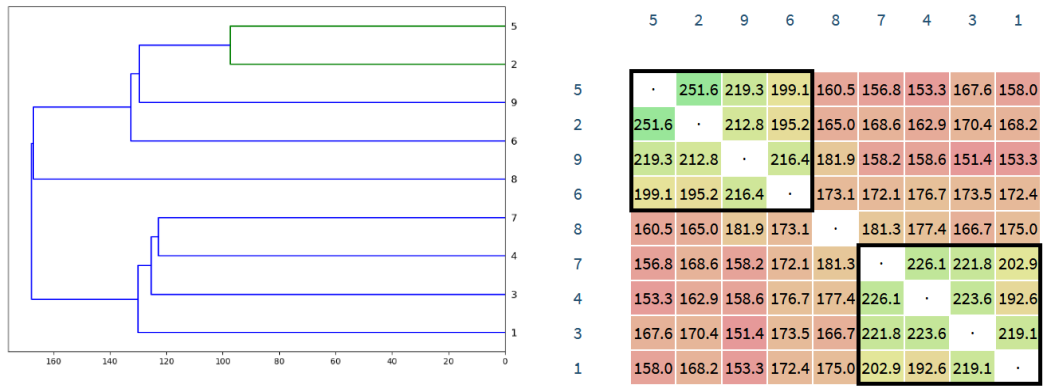

Figure 11: Dendrogram (left) and global similarity matrix (right) for SensoGraph with distances. The matrix is rearranged according to the result of the dendrogram, so that the groups obtained appear as submatrices with similar colors (framed).

Figure 9 shows the consensus graphic obtained, with all the connections between samples, using SensoGraph with distances, as introduced in Section 2.5. Figure 10 shows only the 60% of most relevant connections. Finally, Figure 11 depicts the dendrogram (left) obtained from the global similarity matrix (right). Again, the color code from red to green has been used for the consensus graphics and the similarity matrix.

The consensus map obtained (Figure 9) is similar to those given by the previous methods, the RV coefficient with MFA for the first two dimensions (Figure 5) being 0.7407 and that with SensoGraph with Gabriel (Figure 6) being 0.6390. Both the consensus maps (Figures 5, 6, and 9) and the dendrograms and global similarity matrices (Figures 8 and 11) show the same groups, being slightly more clear for SensoGraph with distances than for the variant with Gabriel (Figures 7 and 10, also Figures 8 and 11).

**3.4 Stability of the results**

As detailed in Section 2.6, bootstrapping was performed to analyze the stability of the results given in the previous subsections by the three methods considered.

Figures 12 to 16 show, for different cases, the evolution of the RV and Mantel coefficients between virtual panels, composed by subsets of consumers randomly drawn with replacement, and the true panel. The vertical axis corresponds to the average RV or Mantel coefficient, incorporating standard deviations as vertical bars, while the horizontal axis corresponds to the number of consumers in the virtual panel. As expected, in all the cases increasing the number of consumers leads to an increase of the RV or Mantel coefficients and a decrease of the standard deviations.

Previous works (Faye et al., 2006, Blancher et al., 2012; Vidal et al., 2014) have considered different values of the RV coefficient as the threshold above which the results are considered to be stable, the most restrictive one being the value 0.95 proposed by Blancher et al. (2012), which is depicted as a red horizontal dashed line. There is no agreement in the literature about which value of the RV coefficient indicates a good agreement, with Vidal et al. (2014) reporting works which consider values that range between 0.65 and 0.95. For the Mantel coefficient there is no standard threshold value as well, although Blancher et al. (2012) observed a strong linear relationship between both types of coefficient, with the Mantel coefficient tending to provide slightly lower values than the RV coefficient.

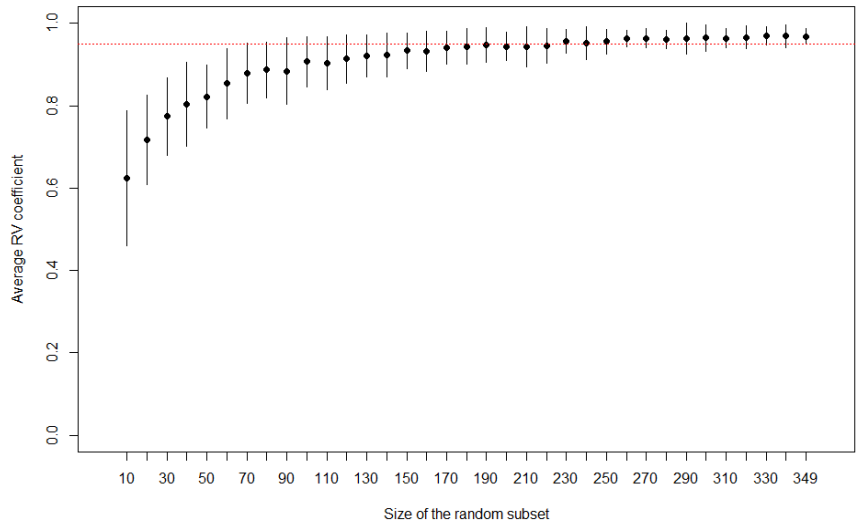

Figure 12: Evolution of the RV coefficient between subsets of consumers randomly drawn with replacement and the whole panel, for the first two dimensions of MFA (Figure 5). The red horizontal dashed line corresponds to a 0.95 threshold.

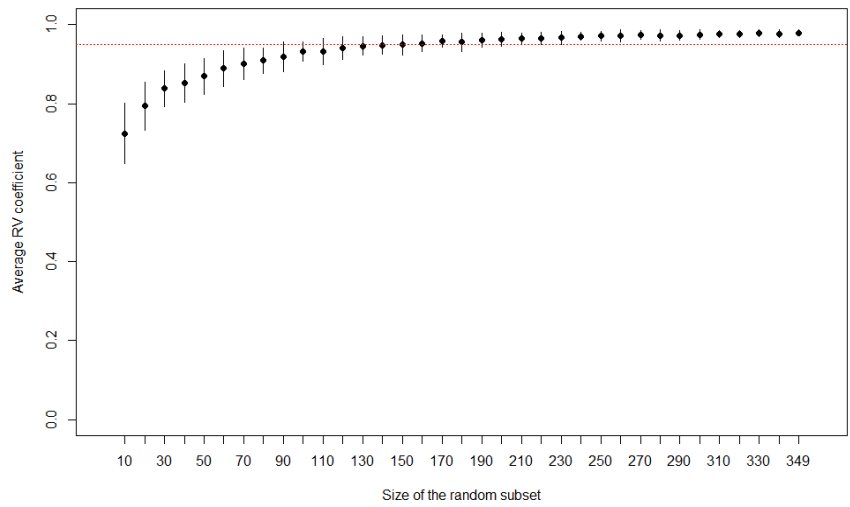

Figure 13: Evolution of the RV coefficient between subsets of consumers randomly drawn with replacement and the whole panel, for the first four dimensions of MFA. The red horizontal dashed line corresponds to a 0.95 threshold.

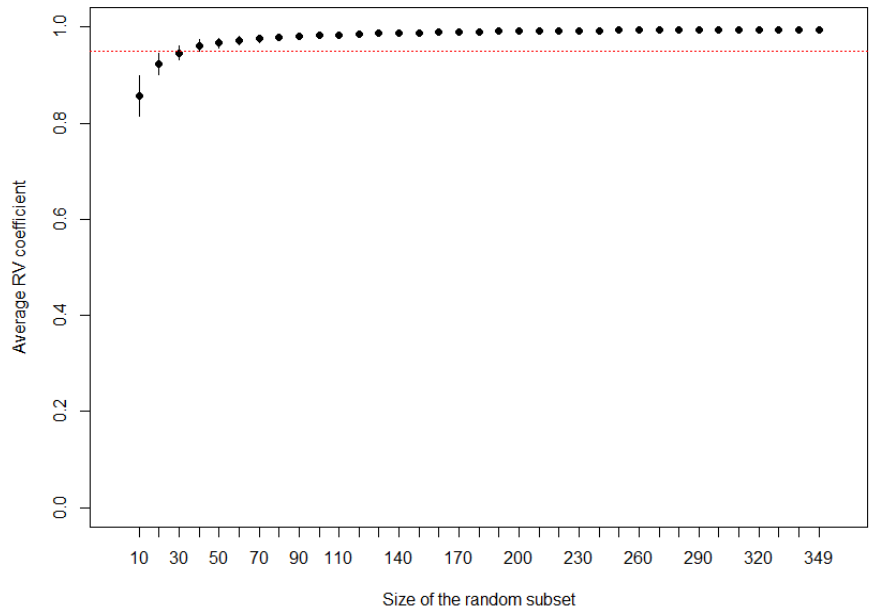

Figure 14: Evolution of the RV coefficient between subsets of consumers randomly drawn with replacement and the whole panel, for the eight dimensions of MFA. The red horizontal dashed line corresponds to a 0.95 threshold.

For the case of MFA, considering the first four dimensions instead of the first two leads to better results, with higher initial RV coefficients and smaller standard deviations. Vidal et al. (2014) observed similar results where, their work considering up to 100 consumers, some panels did not reach the 0.95 RV coefficient suggested by Blancher et al. (2012). The results we obtain show that for this panel of unexperienced consumers with the first two dimensions accounting for only 41.12% of the explained variance, the 0.95 threshold is achieved for around 200 consumers (Figure 12). When considering the first four dimensions, accounting for 65,95% of the explained variance, the 0.95 threshold is achieved for around 150 consumers (Figure 13). Considering the eight dimensions, which obviously account for the 100% of the explained variance, allows to achieve the 0.95 threshold for 40 consumers (Figure 14).

It should be noticed that Næs et al. (2017) analyzed the corpus of 46 publications in Food Quality and Preference and Food Research International dealing with projective mapping until then, concluding that most of the papers considered only the first two dimensions of the MFA, with just a few of them (6 out of 46) going further and using up to the first four dimensions. To the best of our knowledge, no previous work used the full (eight) dimensions of MFA to study the stability of the samples configuration.

As for SensoGraph with Gabriel, the results in Figure 15 show smaller deviations than those for MFA with the first two dimensions (Figure 12), although more consumers are needed to achieve the 0.95 threshold, around 300 consumers. This could be explained because, as observed by Blancher et al. (2012), the Mantel coefficient tends to be slightly smaller than the RV coefficient. Actually, in this case, considering 200 consumers achieves a 0.92 threshold.

It is worth noticing that, although the dimensionality of the 9x9 global similarity matrix is eight (it defines nine points in nine dimensions but the fact that entries *(i,i)* are zero implies that those points do actually lie on an 8-dimensional hyperplane), the global similarity matrix can also be depicted in a single 2-dimensional graphic (Figure 8, right).

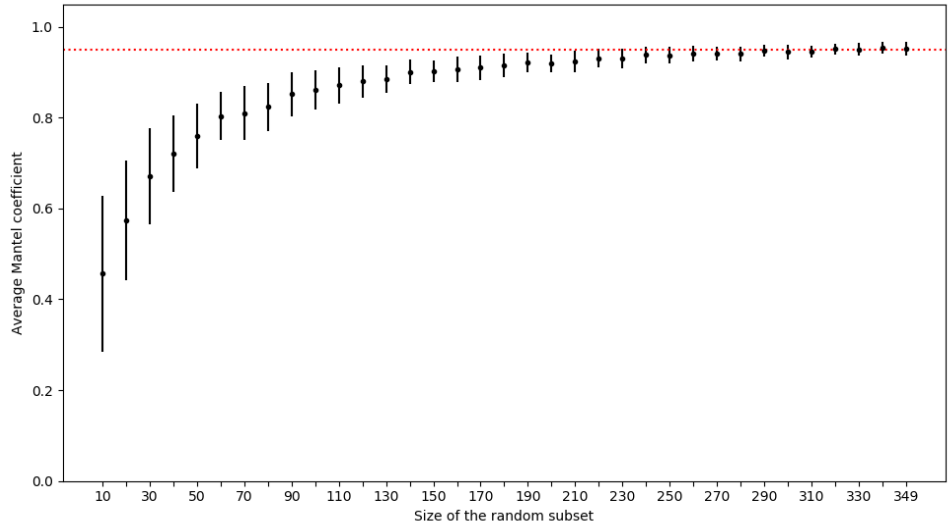

Figure 15: Evolution of the Mantel coefficient between subsets of consumers randomly drawn with replacement and the whole panel, for the SensoGraph with Gabriel global similarity matrix (Figure 8). The red horizontal dashed line corresponds to a 0.95 threshold.

Finally, SensoGraph with distances leads to the results depicted in Figure 16, which show higher Mantel coefficients and smaller standard deviations than the version using Gabriel. In particular, the 0.95 threshold is achieved for around 200 consumers, like for MFA with the first two dimensions (Figure 12), but with smaller deviations. Further studies with different panels and products, as in (Vidal et al., 2014), should be performed in order to confirm this behavior.

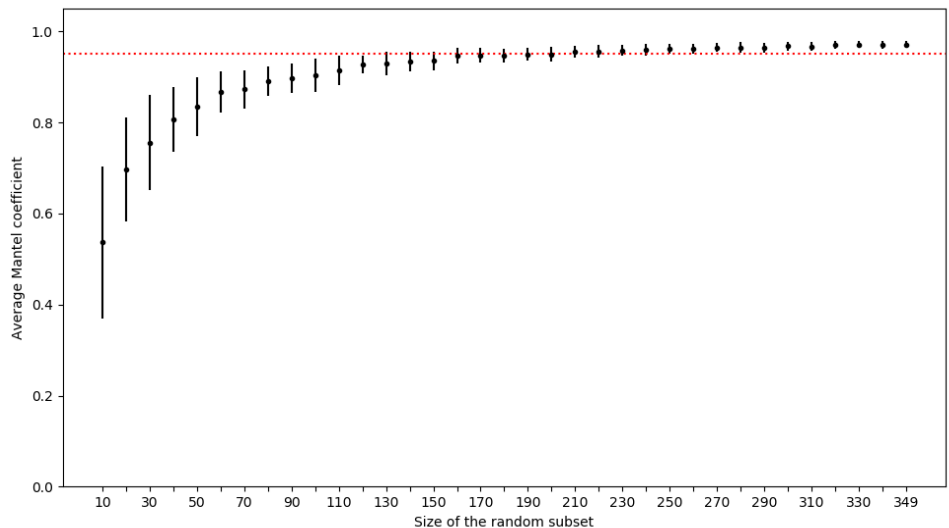

Figure 16: Evolution of the Mantel coefficient between subsets of consumers randomly drawn with replacement and the whole panel, for the SensoGraph with distances global similarity matrix (Figure 11). The red horizontal dashed line corresponds to a 0.95 threshold.

## 4.-CONCLUSIONS

This work used Projective Mapping for the evaluation of commercial chocolate chip cookies by a large number of n=349 unexperienced consumers, analyzing the data with statistical (MFA), geometric (SensoGraph-Gabriel) and mixed (SensoGraph-distances) methods. All of them provided the same groups of samples, with the two blind duplicates being positioned close together. The identification of these duplicates was clearer for the geometric and mixed techniques than for the MFA consensus map, where a different pair of samples appeared as the closest one.

The stability of the results was studied with bootstrapping resampling, randomly drawing 100 subsets of m=10, 20, 30,…, n assessors from the original data set and measuring the agreement with the original panel by the RV coefficient for MFA and the Mantel coefficient for SensoGraph. MFA achieved the highly restrictive RV 0.95 stability threshold for around 200 consumers when using the first two dimensions, for around 150 consumers when considering the first four dimensions, and for 40 consumers when all the eight dimensions are considered. For SensoGraph with Gabriel, the Mantel 0.95 stability threshold was achieved around 300 consumers, while SensoGraph with distances led to values beyond the Mantel 0.95 threshold for around 200 consumers. These values are to be considered taking into account that Mantel coefficients have been previously observed to be slightly smaller than RV coefficients.

Further research, with different panels and products, would be needed in order to confirm these behaviors which suggest that, on one hand, global similarity matrices are useful for Projective Mapping data analysis and, on the other hand, graph drawing techniques provide reliable consensus maps.

## 5.-ACKNOWLEDGEMENTS AND DISCLAIMER

David Orden has been partially supported by Project CCG19/CC-035 (GINSENG) of the University of Alcalá, by Project MTM2017-83750-P of the Spanish Ministry of Science (AEI/FEDER, UE), by Project PID2019-104129GB-I00 of the Spanish Ministry of Science and Innovation, and by H2020-MSCA-RISE project 734922 – CONNECT. Encarnación Fernández-Fernández has been partially supported by Project CCG19/CC-035 (GINSENG) of the University of Alcalá and by Project PID2019-104129GB-I00 of the Spanish Ministry of Science and Innovation. Marino Tejedor-Romero is funded by the University of Alcalá, through the program Ayudas de iniciación en la actividad investigadora. Alejandra Martínez-Moraian is funded by the predoctoral contract PRE2018-085668 of the Spanish Ministry of Science, Innovation, and Universities, and partially supported by Project PID2019-104129GB-I00 of the Spanish Ministry of Science and Innovation. Part of her work on this paper was made during a stay at the University of Alcalá while she was with the Departamento de Matemáticas, Estadística y Computación at the University of Cantabria.

The authors declare to have no conflict of interest regarding any of the commercial brands involved in this work. They also want to thank the anonymous reviewers for their useful comments, which helped to improve the manuscript.

## 6.-REFERENCES